\documentclass{article}

\usepackage[dvipdfmx]{graphicx}

\usepackage{url}

\begin{document}
	
\title{Visualization of the Computation Process of \\a Universal Register Machine}

\author{Shigeru Ninagawa\footnote{Kanazawa Institute of Technology, Japan. E-mail: ninagawa@neptune.kanazawa-it.ac.jp }
	\and Genaro J. Mart{\'i}nez\footnote{Escuela Superior de C\'omputo, Instituto  Polit\'ecnico Nacional, M\'exico.
	Unconventional Computing Laboratory, University of the West of England, Bristol, United Kingdom. E-mail: genaro.martinez@uwe.ac.uk}
}

\date{28 October 2020}

\maketitle

%% 100--150 words
\begin{abstract}
Universal register machine, a formal model of computation, can be emulated on the array of
the Game of Life, a two-dimensional cellular automaton.
We perform spectral analysis on the computation dynamical process of the universal register machine on
the Game of Life.
The array is divided into small sectors and the power spectrum is calculated from the evolution
in each sector. 
The power spectrum can be classified into four categories by its shape; null, white noise,
sharp peaks, and power law.
By representing the shape of power spectrum by a mark, we can visualize the activity
of the sector during the computation process. 
For example, the track of pulse moving between components of the universal register machine
and the position of frequently modified registers can be identified.
This method can expose the functional difference in each region of computing machine.
\end{abstract}

%%\keywords{universal register machine, computation process, Game of Life, functional imaging, spectral analysis.}
	
%%%%%%%%%%%%%%%%%%%%%%%%%%%%%%%%%%%%%%%%%%%%%%%%%%%%%%%%%%%%

\section{Introduction}

A wide variety of formal models of computation capable of supporting effective computability 
have been proposed in the theory of computation.
Although all these models are computationally equivalent in the sense
that they can emulate each other, there is much difference among them.
The most noticeable difference is the `concreteness' of the model.
While partial recursive function is a set of functions and there is no notion of space
and time in its definition, Turing machine (TM) is concrete enough to make as an actual machine.
In analyzing the computation process of a model that works in space and time,
we can make use of several kinds of information accompanied with computation process.
For example, side-channel attack is one of cryptanalysis to break the cryptographic
security by exploiting physical signals such as power consumption~\cite{Koch99} or electromagnetic
radiation~\cite{Gan01} of cryptographic module.

The innovation in visualization technology have highly promoted science in history, e.g. microscope
in biology or telescope in astronomy.
Nowadays neuroscience has achieved considerable progress by means of functional neuroimaging
such as PET (positron emission tomography) or fMRI (functional magnetic resonance imaging).
These visualization technologies can visualize the local activity of living brain
by measuring blood flow or metabolic processes associated with neuronal activity.
Visualization of computation process might provide a novel insight into the nature of
computation.

In this research we deal with cellular automata (CA) as a model of computation
because it is easier to acquire and analyze the data on computing processes.
We particularly employ universal register machine (URM)~\cite{chapman_web} constructed
in the array of the Game of Life (LIFE)~\cite{BCG82}.

In one-dimensional CAs, space-time pattern~\cite{Wol83} is widely used to display the temporal
behaviour of CA and furthermore several filtering methods~\cite{Han92}, \cite{Wue99}, \cite{Hel04},
\cite{Sha06}, \cite{Liz08} have been developed to extract essential structures from
space-time pattern.
In two-dimensional CAs, three-dimensional display has been used  \cite{Mar21}.

In this paper we try to display the functional difference instead of showing bare state of cell
in two-dimensional CAs.
%We use power spectrum as a measure to display that is integrated in the time domain
The pattern of the URM on the array of LIFE is stationary as a whole but
it fluctuates in minute scales. So the fluctuation in an area indicates the activity of the area.
We use power spectrum as a measure of activity of the area because it is integrated from the fluctuation.

This paper is organized as follows. We explain the URM constructed on the array of LIFE in section 2.
The visualization method based on spectral analysis is given in section 3. We discuss the meaning
of the results and the futures plans in section 4.

\section{Model of Computation}

In this section, we explain the mechanism of the URM and its implementation on the array of LIFE.

\subsection{Universal Register Machine}
Register machine \cite{Minsky67}, \cite{Dennet08} is a formal model of computation with
a finite (or infinite) set of registers. We number them consecutively from zero.
Each register can hold an indefinitely large non-negative integer.
The register machine employed in this article is supposed to have an instruction set
listed in Table \ref{table:InstSet}.

\begin{table}[htb]
\caption{Instruction set of the register machine.}
\begin{center}
\begin{tabular}{llll}
mnemonic & Operand 1 & Operand 2 & Operand 3  \\
\hline
INC  &  n  & PassAdr &                \\
DEC  & n  & PassAdr  & FailAdr   \\
HALT &    &               &            \\
\hline
\end{tabular}
\label{table:InstSet}
\end{center}
\end{table}

The {\it INC} instruction increments the content of register $n$ and jumps to address {\it PassAdr}.
The {\it DEC} instruction decrements from the content of register $n$ if it is greater
than 0 and jumps to address {\it PassAdr}, otherwise it does nothing but jump to
address {\it FailAdr}.
The {\it HALT} instruction halts the program.
For example, the program listed in Table \ref{table:Program} adds the content of register 1
to register 0 and resets register 1.

\begin{table}[htb]
\caption{Sample program of the register machine.}
\begin{center}
\begin{tabular}{lllll}
address & mnemonic & Operand1 & Operand2 & Operand3 \\
\hline
0 & DEC  &  1  & 1 & 2   \\
1 & INC  &  0  &  0 &    \\
2 & HALT &    &     &      \\
\hline
\end{tabular}
\label{table:Program}
\end{center}
\end{table}

Register machine performs a fixed task according to a prestored program, whereas
URM can emulate the behavior of any register machine.
Let $M$ be a register machine and $U$ be a URM.
By encoding both the program and the contents of the registers of $M$ into the registers of $U$,
$U$ can emulate the behavior of $M$.

The URM adopted in this article has 12 registers.
Let $r_i$ (i = 0, 1, 2, $\cdots$) denote the content of register $i$ of $M$ and
$R_j$ (j = 0, 1, $\cdots$, 11) the content of register $j$ of $U$.
We encode the values $r_i$ into $R_0$ by G\"{o}del numbering as follows;
\begin{equation}
R_{0} = P(1)^{r_0} \times P(2)^{r_1} \times \cdots,
\end{equation}
where $P(n)$ denotes the $n$ the prime number ($P(1)$=2, $P(2)$=3, $\cdots$).
Given $r_0$ = $r_1$ = 1 and $r_i$ = 0 ($i>1$), we have $R_0$ = $2^1 \times 3^1 $ = 6.

\begin{figure}
\begin{center}
\includegraphics[width=0.4\linewidth]{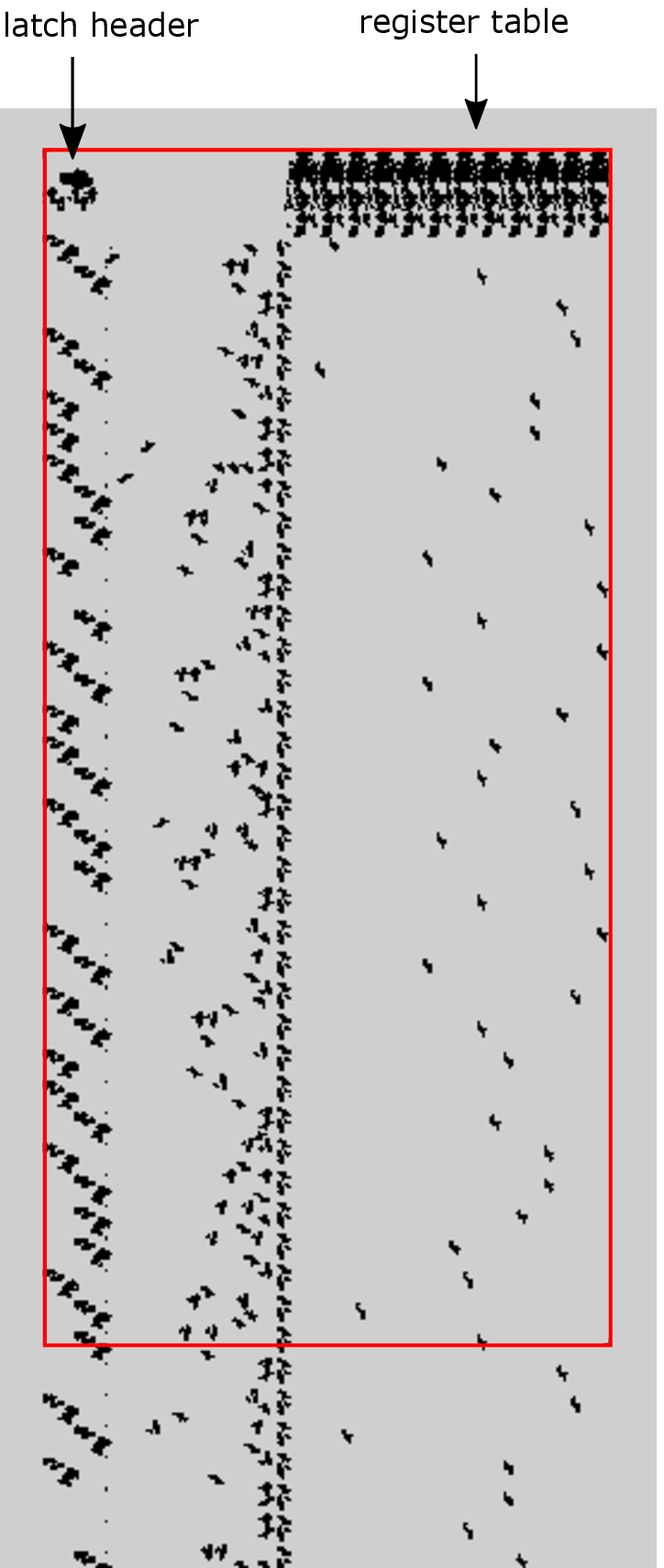}
%%\hspace{15mm}
\includegraphics[width=0.4\linewidth]{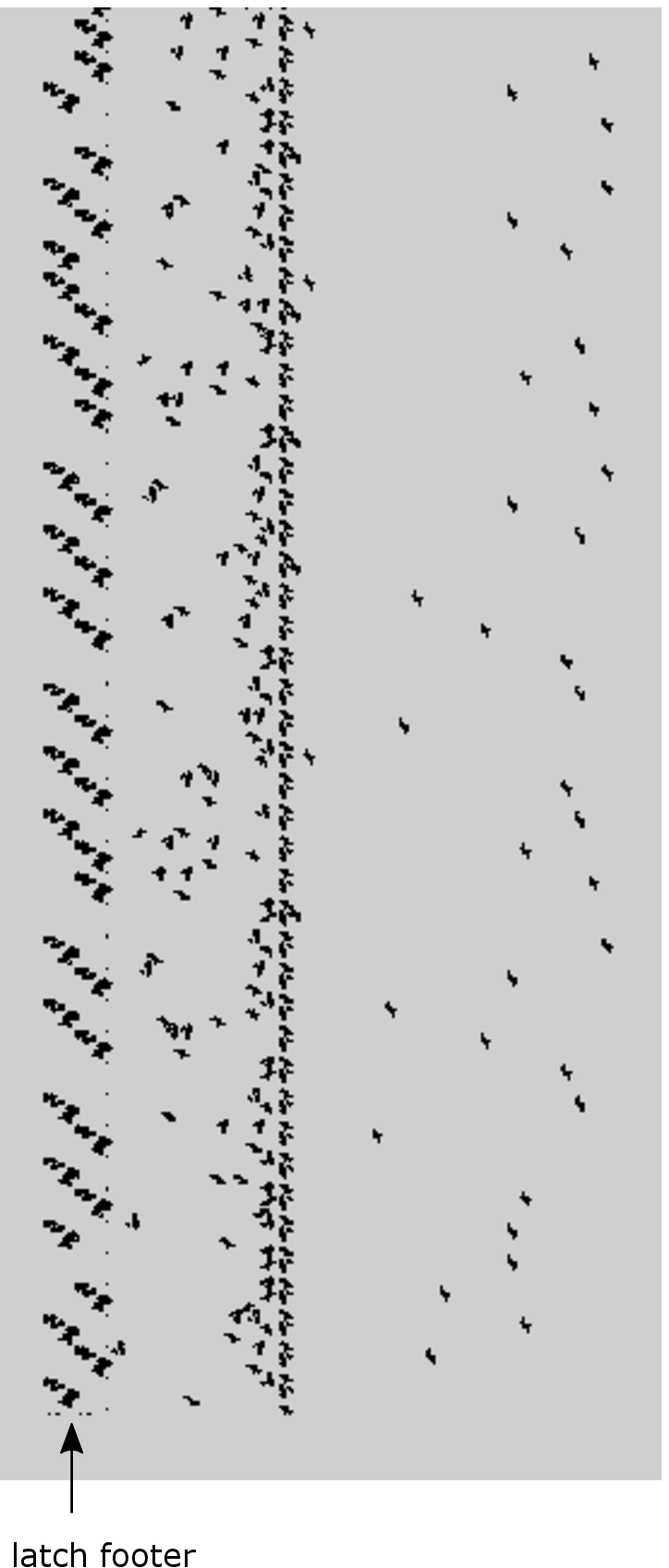}
\end{center}
\caption{Initial configuration to emulate the URM on LIFE.
The upper (lower) half of the configuration is on the left (right).  
The gray and black squares are the cells with state zero and one.
The rectangle drawn in a solid line represents the area in which power spectra are calculated.}
\label{fig:InitConf}
\end{figure}

We encode a series of operators of the program of $M$ into $R_1$ by assigning zero for {\it HALT},
one for {\it INC}, and two for {\it DEC}.
Let $I_i \in \{0,1,2\}$ be a number corresponding to the operator of the program of $M$ in address
$i$  (i = 0, 1, 2, $\cdots$). We set $R_1$ as follows;
\begin{equation}
R_{1} = P(1)^{I_0} \times P(2)^{I_1} \times \cdots.
\end{equation}
As for the program listed in Table \ref{table:Program}, we have $R_1$ = $2^2 \times 3^1 \times 5^0$ = 12.

We encode a series of the first operands of the program of $M$ into $R_2$.
Let $q_i$  (i = 0, 1, 2, $\cdots$) denote the first operand of the program of $M$ in address $i$.
If there is no operand in an instruction such as {\it HALT}, $q_i$ is equal to zero.
We set $R_2$ as follows;
\begin{equation}
R_{2} = P(1)^{P(q_0+1)-2} \times P(2)^{P(q_1+1)-2} \times \cdots.
\label{eq:encode}
\end{equation}
We have $q_0$ = 1, $q_i$ = 0 ($i > 0$) in the program in Table \ref{table:Program}, 
so \ref{eq:encode} gives $R_2$ = $ 2^{P(2)-2} \times 3^{P(1)-2} \times 5^{P(1)-2} $ = 2.

As for the second and third operand, we encode them into $R_3$ and $R_4$ in the
same way as (\ref{eq:encode}). So we have $R_3$ = 2, $R_4$ = $ 2^{P(3)-2} \times 3^{P(1)-2} \times 5^{P(1)-2} $ = 8
for the program in Table \ref{table:Program}.
We unconditionally set $R_5$ = $P(1)$ = 2 and $R_6$ = $R_7$ = $\cdots$ = $R_{11}$ = 0.

\subsection{Implementation}

LIFE is a two-dimensional CA that has computational universality.
Let $s_{x,y}(t) \in \{ 0, 1 \}$ be the state of the cell $(x, y)$ at time step $t$
and $n_{x,y}(t)$ denote the number of state one cells among surrounding eight cells
of the cell $(x, y)$ at time step $t$.
The evolution of each cell is governed by the transition rule,
\begin{equation}
s_{x,y}(t+1) = F(s_{x,y}(t),n_{x,y}(t)),
\end{equation}
where $F$ is a transition function defined by
\begin{eqnarray}
F(0,3) = F(1,2) = F(1,3) = 1, \nonumber \\
otherwise \hspace{1cm} F = 0.
\label{eq:rule}
\end{eqnarray}

Figure~\ref{fig:InitConf} shows the initial configuration of LIFE that can emulate
the above-mentioned URM. 
The gray and black square represents the cell with state zero and one respectively.
%This pattern file can be downloaded from the website~\cite{rendell_UTM} and it can be run on Golly,
%a simulator of the Game of Life~\cite{Golly}.
The pattern of the URM spans about 19,000 cells in height and 3,900 cells in width and it takes
about 32,586,000 time steps to halt.
The URM we deal with in this article emulates the register machine that performs the program
described in Table \ref{table:Program} with $r_0$ = $r_1$ = 1 and $r_i$ = 0 ($i>1$).
The transition of each register of the URM is shown in Table \ref{table:EvoReg} in which
$t$ denotes the generation of LIFE emulating the URM.

\begin{table}[t]
\caption{Transition of the contents of register 0$\sim$11 of the URM
emulating the program in Table \ref{table:Program}. $t$ denotes the generation in the LIFE on which the URM runs.}
\begin{center}
\begin{tabular}{r|llllllllllll}
t          & 0  &  1    & 2  & 3 & 4 & 5 & 6 & 7 & 8 & 9 & 10 & 11 \\
\hline
0        & 6  &  12  & 2  & 2 & 8 & 2 & 0 & 0 & 0 & 0 & 0   &   0  \\
5,300   & 6  &  11  & 2  & 2 & 8 & 2 & 0 & 0 & 0 & 0 & 0   &   0  \\
20,154 & 6  &  11  & 2  & 2 & 8 & 2 & 0 & 1 & 0 & 0 & 0   &   0  \\
20,700 & 6  &  10  & 2  & 2 & 8 & 2 & 0 & 1 & 0 & 0 & 0   &   0  \\
22,909 & 6  &  10  & 2  & 2 & 8 & 2 & 0 & 1 & 0 & 0 & 1   &   0  \\
35,967 & 6  &  10  & 2  & 2 & 8 & 2 & 0 & 2 & 0 & 0 & 1   &   0  \\
36,500 & 6  &   9   & 2  & 2 & 8 & 2 & 0 & 2 & 0 & 0 & 1   &   0  \\
38,719 & 6  &   9   & 2  & 2 & 8 & 2 & 0 & 2 & 0 & 0 & 2   &   0  \\
51,804 & 6  &   9   & 2  & 2 & 8 & 2 & 0 & 3 & 0 & 0 & 2   &   0  \\
52,373 & 6  &   8   & 2  & 2 & 8 & 2 & 0 & 3 & 0 & 0 & 2   &   0  \\
54,562 & 6  &   8   & 2  & 2 & 8 & 2 & 0 & 3 & 0 & 0 & 3   &   0  \\
\multicolumn{1}{c|}{$:$} & $:$ & $:$ & $:$ & $:$ & $:$ & $:$ & $:$ & $:$ & $:$ & $:$ & $:$ & $:$ \\
32,586,211 & 4  &   12   & 2  & 2 & 8 & 5 & 0 & 0 & 0 & 0 & 0   &   0  \\
\end{tabular}
\label{table:EvoReg}
\end{center}
\end{table}

Taking a broad view of the URM on LIFE, it is composed of five parts.
At the top row from left to right, there exist {\it latch header} and {\it register table}.
The {\it latch header} turns a pulse coming from a register in a downward direction and
the {\it register table} contains 12 registers.
In the middle row from left to right, there are {\it table area} and {\it operation table}.
The most part of the configuration is occupied by these two area.
The {\it table area} consists of six kinds of tables and the {\it operation table} is studded
with {\it operation merges}.
At the bottom row there is {\it latch footer} that terminates a pulse coming through the
{\it table area}.
The signal transmission between these components is conveyed by pulse called light weight space
ship (LWSS).% and its evolution is shown in Figure~\ref{fig:LWSS}.
A pulse repeatedly circulates through {\it table area}, {\it operation table},
{\it register table}, and {\it latch header} and finally vanishes at {\it latch footer}. 

%\begin{figure}[htb]
%\begin{center}
%\includegraphics[width=0.7\linewidth,clip]{figs/LWSS.eps}
%\end{center}
%\caption{Evolution of a Light Weight Space Ship (LWSS).
%		The white and black square represents the cell with state zero and one respectively.}
%\label{fig:LWSS}
%\end{figure}

Next let us show you the computation process of the URM in detail.
The computation process starts when a pulse called {\it execution pulse} goes left
from the top of {\it table area} to {\it operation table} that turns the pulse in a upward
direction and the pulse reaches one of the registers.

\begin{figure}
\begin{center}
\includegraphics{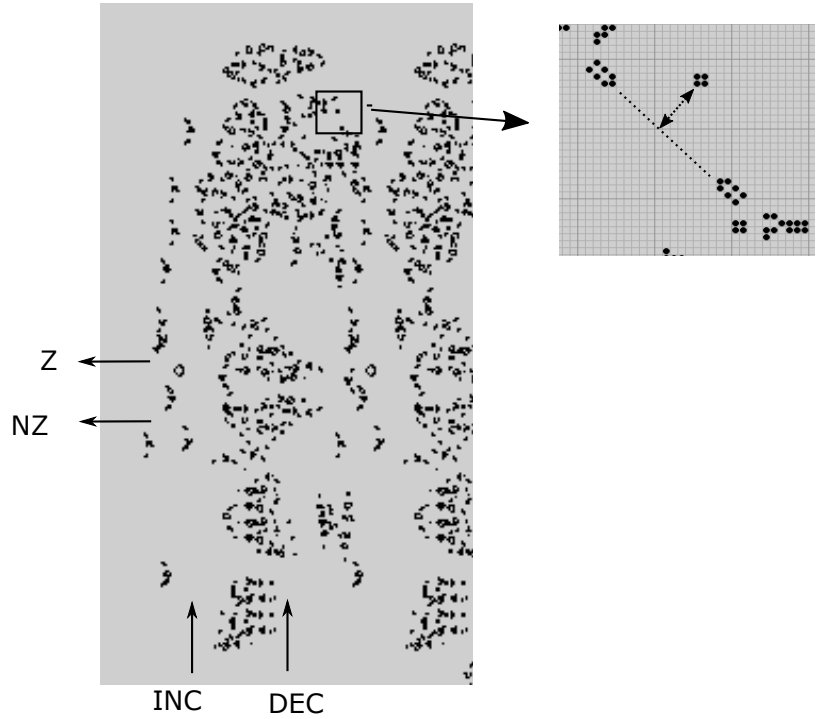}
\end{center}
\caption{Left: snapshot of a single register. Right: Enlarged view of the area surrounded
	by a solid line.}
\label{fig:register}
\end{figure}

Figure~\ref{fig:register} shows the snapshot of a register and 
and its enlarged view of the area surrounded by a solid line.
The content of the register is held as the length of a bidirectional arrow
in the right of Figure~\ref{fig:register}.
Register has two input gates labelled `INC' and `DEC' at the bottom of
Figure~\ref{fig:register}.
If a pulse enters INC (DEC) gate,  {\it INC} ({\it DEC}) instruction is carried out on the register.
If the content of the register becomes zero as a result of execution of the instruction,
{\it Z pulse} is emitted from Z gate otherwise {\it NZ pulse} is emitted from NZ gate.
{\it Z pulse} and {\it NZ pulse} leave for {\it latch header}.

\begin{figure}
\begin{center}
\includegraphics[width=0.8\linewidth]{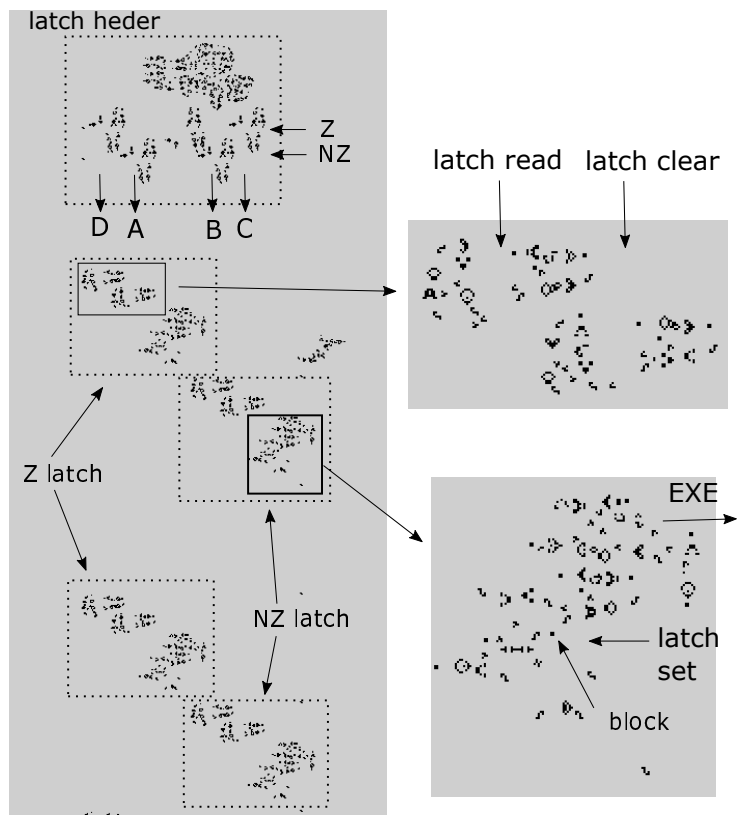}
\end{center}
\caption{Left: pattern of {\it latch header} and {\it Z} and {\it NZ latches}. Upper right: Enlarged view of the left part
of latch surrounded by a solid line. Lower right:  Enlarged view of the right part of latch surrounded by solid line.}
\label{fig:latch}
\end{figure}

{\it Latch header} has two input gates labelled `Z' or `NZ' and four output gates labelled
`A', `B', `C', and `D' as shown in the left of Figure~\ref{fig:latch}.
If an {\it NZ pulse} enters the {\it NZ gate}, a {\it latch clear pulse}
is emitted from output gate A and a {\it latch read pulse} from output gate B.
Similarly if a {\it Z pulse} enters the {\it Z gate}, a {\it latch clear pulse} is emitted from output gate C and a {\it latch read pulse} from output gate D.

The areas surrounded by dotted line below the latch header are latches.
The latch has three input gates called "latch read", "latch clear", and "latch set" and
has three output gates "latch read", "latch clear", and "execution".
Two of the outputs, "latch read" and "latch clear" are the duplication of each inputs.
The latches located in the left column are called {\it Z latches} and {\it NZ latches}
in the right.

The upper right of Figure~\ref{fig:latch} shows an enlarged view of the upper left part of a Z latch.
{\it Latch read pulse} and  {\it latch clear pulse} enter the Z latch from above following the arrows with respective labels.
The lower right of Figure~\ref{fig:latch} is an enlarged view of the lower right part of an NZ latch.
If there is a pattern pointed out with a label `{\it block}', the latch is being set, otherwise it is being clear.

The latch is initially being the clear state. If a {\it latch set pulse}, that is mentioned below, enters a latch from the right
following the arrows with label `latch set', the latch becomes the set state.
If a {\it latch set pulse} enter the latch being the set state, it becomes the clear state.

When a {\it latch clear pulse} passes downward through a latch, the latch becomes the clear
state if it is being set, otherwise the latch keeps being clear. 
When a {\it latch read pulse} passes downward through a latch,  the latch becomes the clear
state if it is being set and an {\it execution pulse} is emitted along the arrow labelled
`EXE' in the  lower right of Fig.~\ref{fig:latch}.
The {\it latch read pulse} does not change the latch being clear.
{\it Latch clear pulse} and {\it latch read pulse} are terminated in colliding with
{\it latch footer}.
In this computation process, {\it latch read pulse} is terminated at about $t=32,586,000$
and the whole pattern ends up in periodic behavior.

\begin{figure}[t]
\begin{center}
\includegraphics[width=0.86\columnwidth]{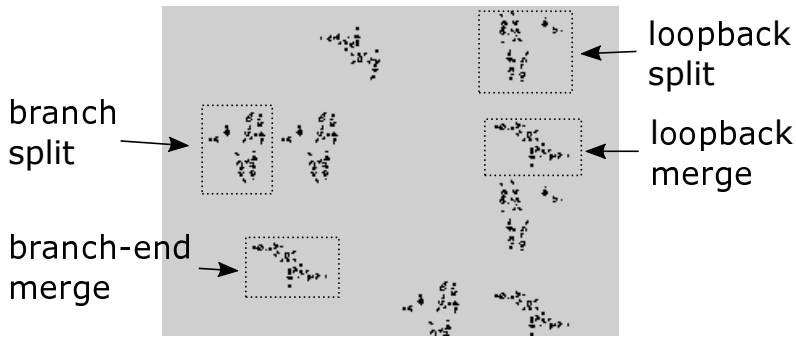}
\end{center}
\caption{Some structures in the {\it table area.}}
\label{fig:Tables}
\end{figure}

{\it Table area} is composed of {\it latch table}, {\it loopback pulse eater table}, {\it branch table}, {\it loopback table}, and {\it NOP/HALT eater table} starting from the left.
Figure \ref{fig:Tables} shows several structures located in the {\it table area}.

While {\it execution pulse} emitted from a {\it Z/NZ latch} passes through the {\it table area},
{\it latch set pulse} is emitted from the {\it table area} and proceeds to {\it Z/NZ latch}.
When the {\it execution pulse} passes through a {\it loopback split},
a {\it loopback pulse} is emitted downward.
The {\it loopback pulse} turns west at a {\it loopback merge}.
When the {\it loopback pulse} passes through a {\it branch split}, a {\it branch pulse} is emitted
upward or downward. The {\it loopback pulse} keeps going west and vanishes in a collision with {\it loopback pulse eater},
which is not displayed in Fig.~\ref{fig:Tables}.
The {\it branch pulse} turns west when it collides with {\it branch-end merge} and becomes a {\it latch set pulse}.

\section{Visualization Method}

The URM constructed on LIFE can perform computation utilizing various patterns observed
in LIFE.
They are classified in three categories; stationary, periodic and propagating patterns.
All these patterns are located properly for close cooperation with others.
Some areas vary frequently and others do not. That implies that there is
a regional difference in behaviour.
So, we employ spectral analysis to study the regional difference of behavior of the URM 
on LIFE.
%Thus, we perform spectral analysis to investigate the behavior of cells of the URM during the
%computation process and utilize the power spectra to visualize the behavior of the URM.

\subsection{Spectral Analysis}
We perform spectral analysis of the computation process of URM  on LIFE to detect 
the regional activity of the array.
The discrete Fourier transform of a time series of states $s_{x,y}(t)$ for
$t = 0, 1, \cdots, T-1$ is given by
\begin{equation}
\hat{s}_{x,y}(f) = \frac{1}{T}\sum_{t=0}^{T-1}s_{x,y}(t){\exp}
(-i\frac{2{\pi}tf}{T}).
\label{eq:DFT}
\end{equation}
We define the power as
\begin{equation}
S(f) = \frac{1}{N} \sum_{x,y} |\hat{s}_{x,y}(f)|^2,
\label{eq:POWER}
\end{equation}
where the summation is taken in $N$ cells in consideration.
In this research, we divide the area into squares with $50 \times 50$ cells,
so the summation is taken in a 2,500 cells.
The period of the component at a frequency $f$ in a power spectrum is given by $T/f$.

It is known that the complicated behavior such as frequent modification in a cell of tape of TM
accompanies power law in power spectrum \cite{Nina19}.
This result implies that the power law in power spectrum can be an indicator of actively
changing area.
We, therefore, estimate the exponent $\beta$ at low frequencies from the least square fitting of power spectrum by
\begin{equation}
\ln S \approx \alpha + \beta \ln f,
\label{eq:LSF}
\end{equation}

\begin{figure}
\begin{center}
\includegraphics[width=0.5\linewidth]{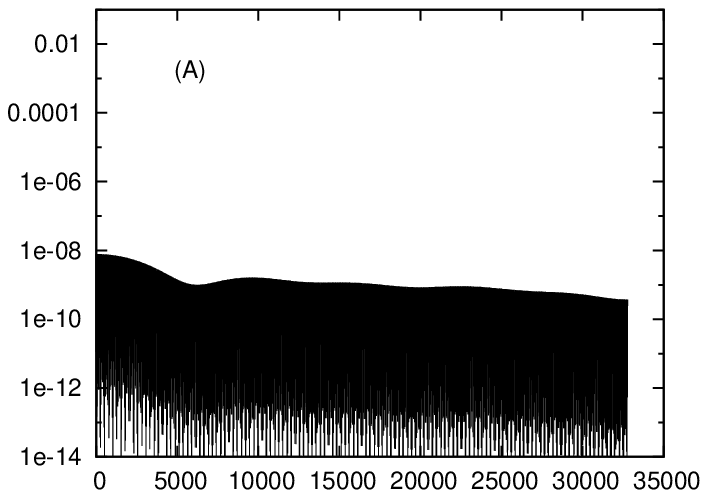}
\includegraphics[width=0.5\linewidth]{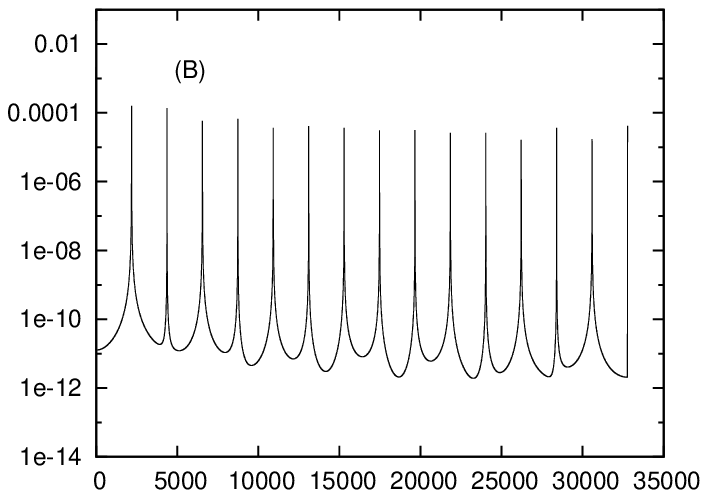}
\includegraphics[width=0.5\linewidth]{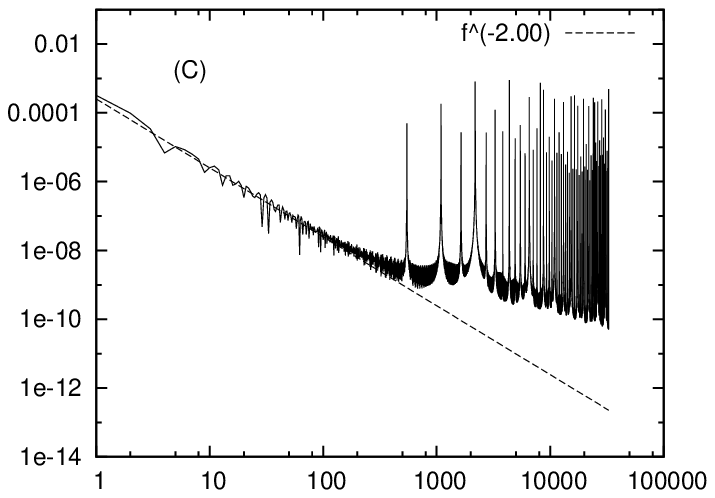}
\end{center}
\caption{Typical examples of power spectra observed in some sectors.
The $x$-axis is the frequency $f$, $y$-axis is power $S$.
The top two are plotted on a semilogarithmic scale and the bottom on a logarithmic scale.
The broken line in the bottom represents the fitting of the power spectrum
from $f=1$ to $f=100$ by $\ln S \approx \alpha + \beta \ln f$ with $\beta = -2.00$.}
\label{fig:PSsec}
\end{figure}

If a power spectrum possesses power law, the degree of the deviation from power law
can be an important measure.
So, we calculate the residual sum of squares $\sigma^2$ defined by
\begin{equation}
\sigma^2=\frac{1}{f_b}\sum_{f=1}^{f_u}{(\ln S - \alpha - \beta \ln f )^2}.
\label{eq:residual}
\end{equation}
%as a measure to detect the derivation of power spectrum from power law.

In this research we set $T$=65,536 and $f_u$ = 100 and calculate the power spectra
on the area with $8,000 \times 3,800$ = 30,400,000 cells surrounded by a solid
line in Figure~\ref{fig:InitConf}.
We divide the area into 160 $\times$ 76 = 12,160 sectors to investigate the regional difference
of power spectra.
Each sector consists of $50 \times 50$ cells.
%If the power spectrum has a large value of residual sum of squares $\sigma^2$, it is not regarded as fitting to power law.
We consider only the power spectrum with $\beta \leq -0.2$ and $\sigma^2 \leq1.5$
as power law.

\subsection{Results}
Most of the sectors exhibit trivial power spectrum in which all components are zero.
Let us call this kind of power spectrum `null' in this article.
The null power spectrum is observed in 9,523 out of 12,160 sectors.
Another trivial type of power spectrum has only DC ($f=0$) component and
this type is observed in 60 out of 12,160 sectors.
Both null and dc-only power spectra are observed in sectors in which there is no change
during the observation time steps. Null power spectra are observed in sectors in which all cells
are in state zero and the sectors with dc-only power spectrum includes stationary patterns. 

\begin{figure}
\begin{center}
\includegraphics[width=0.6\linewidth]{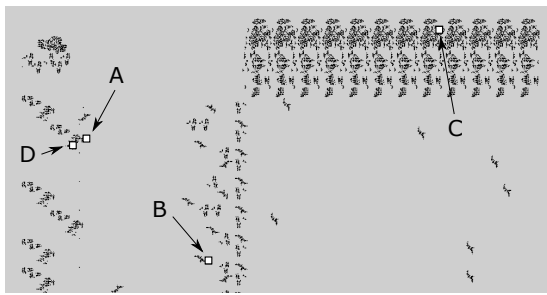}
\end{center}
\caption{Enlarged view of the top part of Figure~\ref{fig:InitConf}. The white squares marked with `A', `B', `C', and `D'
are the sectors where the respective power spectra in Figure~\ref{fig:PSsec} and Figure~\ref{fig:PSsecD} are calculated .}
\label{fig:PSarea}
\end{figure}

Other sectors exhibit distinctive power spectra according to its behavior.
Typical nontrivial examples of power spectra are shown in Figure~\ref{fig:PSsec} (A), (B) and (C) and 
their corresponding positions of sectors are indicated by white squares in Figure~\ref{fig:PSarea}.

The power spectrum in Figure~\ref{fig:PSsec} (A) exhibits white noise.
This kind of power spectrum is observed in 688 out of 12,160 sectors.
%The sectors with white noise are depicted by a white square in Fig.\ref{fig:sectype}.
Since the sectors that exhibit white noise coincide with a pathway of pulse, we can guess
the white noise is caused by the rarely occurred passing of pulses.

\begin{figure}
\begin{center}
\includegraphics[width=0.8\linewidth]{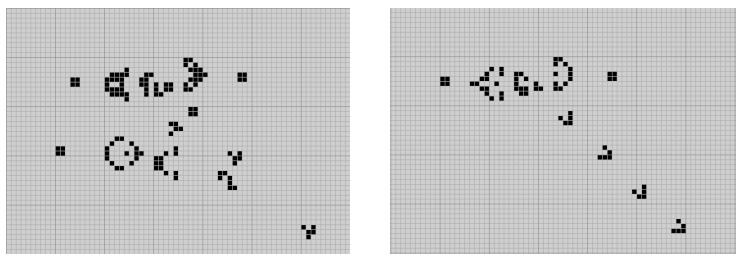}
\end{center}
\caption{Glider gun with period 60 (top) and period 30 (bottom). Gliders are periodically emitted towards the lower right.}
\label{fig:P60gun}
\end{figure}

The power spectrum in Figure~\ref{fig:PSsec} (B) has several peaks.
The sectors in which this type of power spectrum is observed are located on
an oscillator called "glider gun" that can periodically emit propagating pattern called `glider'.
The left of Figure~\ref{fig:P60gun} shows a snapshot of glider gun with period 30 and this oscillator
causes the fundamental frequency with $f = 2,185$ in the power spectrum in
Figure~\ref{fig:PSsec} (B). Another frequently observed glider gun has period 60 and its snapshot is
shown in the right of Figure~\ref{fig:P60gun} and it causes the peak at $f = 1,092$ in other power spectra.
%The gray squares in Fig.\ref{fig:sectype} are the sectors with power spectrum with sharp peaks.
This kind of power spectrum is observed in 1,885 out of 12,160 sectors.

\begin{figure}
\begin{center}
\includegraphics[width=0.4\linewidth]{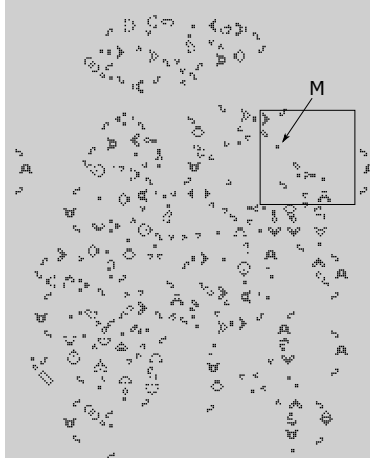}
\end{center}
\caption{Enlarged view of the area around sector C in Figure~\ref{fig:PSarea}.
The square drawn by a solid line is sector C. The block marked with `M' holds the value in the register.}
\label{fig:reg7}
\end{figure}

The power spectrum in Figure~\ref{fig:PSsec} (C) is characterized by power law at low frequencies.
Figure~\ref{fig:reg7} is an enlarged view of the area around the sector C in which the power
spectrum Figure~\ref{fig:PSsec} (C) was observed. The square drawn by a solid line represents sector C in Figure~\ref{fig:PSarea}
and it contains a block pointed out with 'M'. This block 'M' holds the value (zero at the moment) in register 7.
A pair of gliders flying into this sector moves the block 'M' diagonally as the content of the register changes.
%The sectors that exhibit power law are depicted as a black square in Fig. \ref{fig:sectype}.
Power-law type power spectrum is observed in 4 out of 12,160 sectors.

\begin{figure}
\begin{center}
\includegraphics[width=0.7\linewidth]{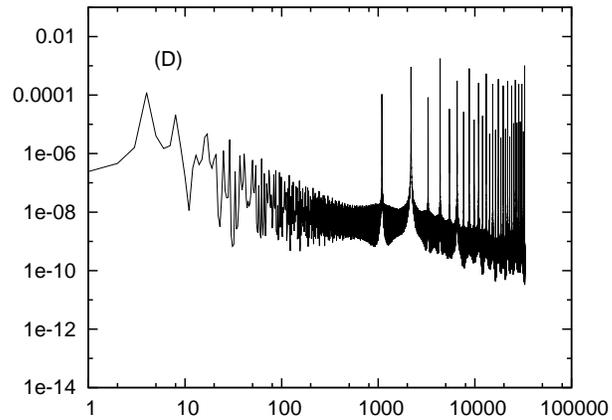}
\end{center}
\caption{Power spectrum in the sector `D' in Figure~\ref{fig:PSarea}.
The $x$-axis is the frequency $f$, $y$-axis is power $S$.
The residual sum of squares $\sigma^2$ in the range between $f=1$ and $f=100$ is 4.19.}
\label{fig:PSsecD}
\end{figure}

The power spectrum in Figure~\ref{fig:PSsecD} is shown as an example of deviation from power law.
This power spectrum seems to exhibit power law at first glance.
But the residual sum of squares $\sigma^2$ is about 4.19.
So this is not considered to be power law according to the criteria $\sigma^2 \leq1.5$ adopted in this
article. 
The position of this sector is depicted in the white square pointed out with `D' in Figure~\ref{fig:PSarea}.
In this sector a normal glider arrives and leaves four times and also an LWSS arrives four times during
65,536 time steps.
Because this sector is slightly active, its power spectrum closely resembles power law.

\begin{figure}
\begin{center}
\includegraphics[width=0.4\linewidth]{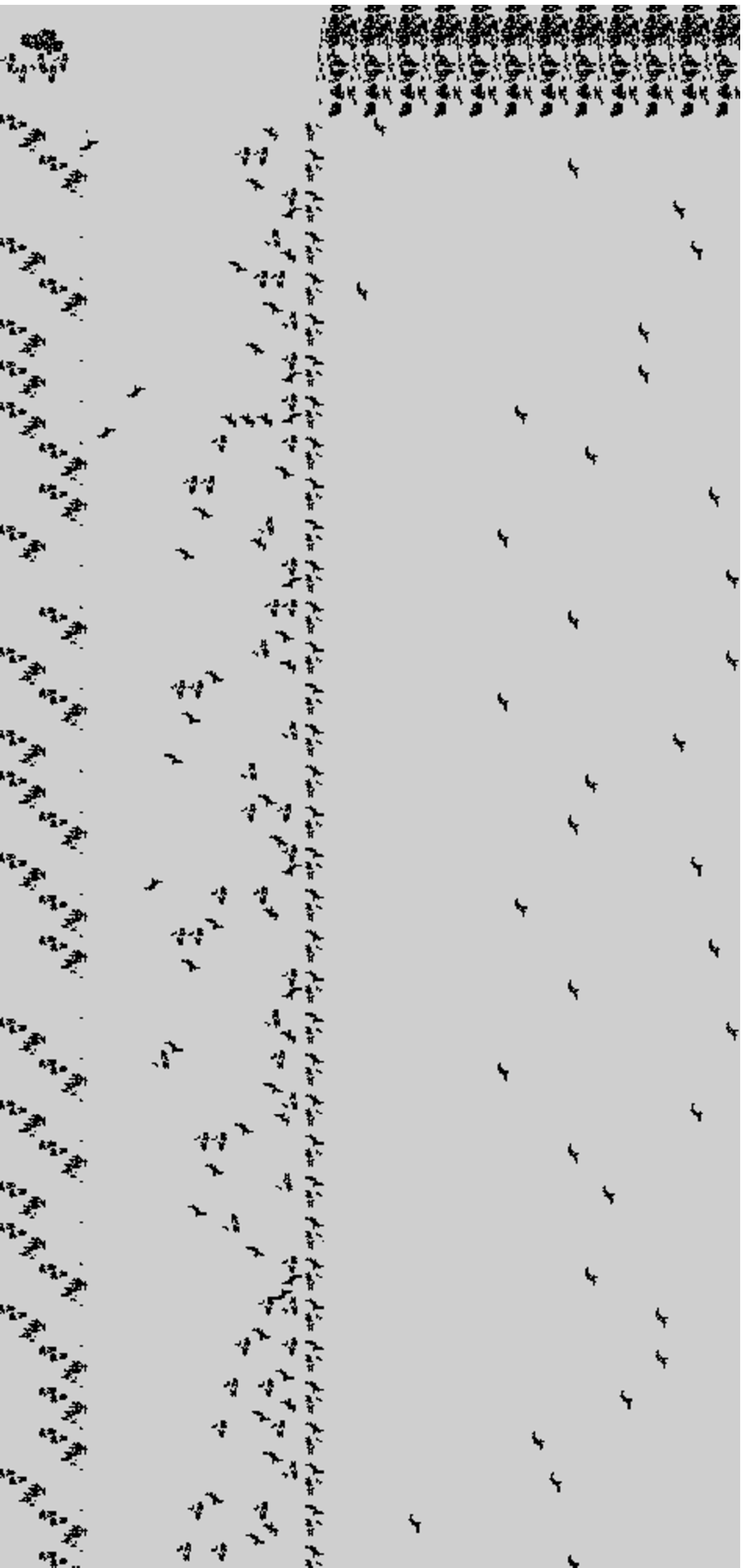}
%%\hspace{3mm}
\includegraphics[width=0.4\linewidth]{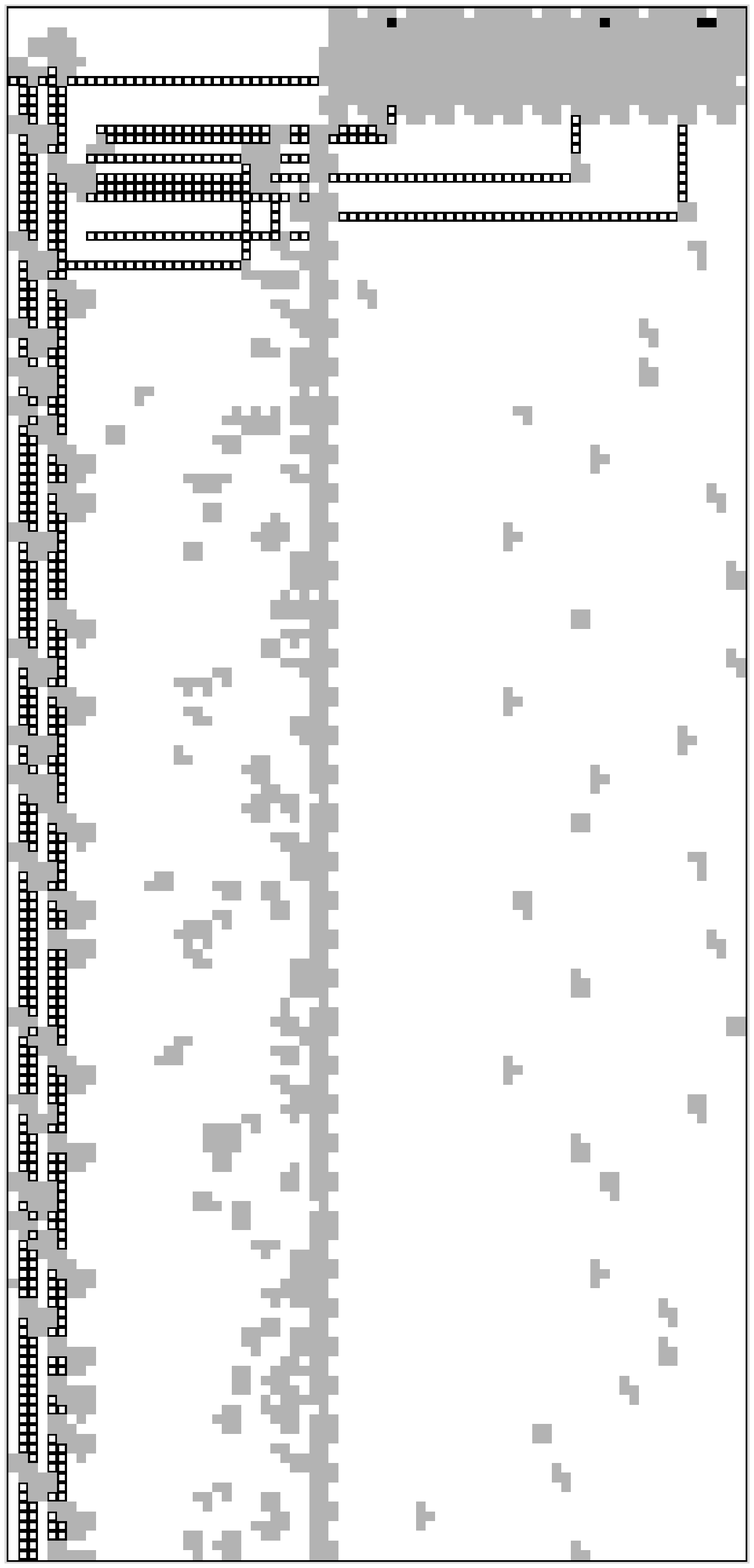}
\end{center}
\caption{Left: Enlarged view of the area surrounded by a solid line in Figure~\ref{fig:InitConf}.
%This figure is located here for the purpose of the comparison with Figure~\ref{fig:sectype}.
Right: Picture drawn by assigning each sector a mark according to the shape of power spectrum.
%The area is the same as Fig. \ref{fig:enlargedConf}.
Blank: null power, white square: white noise, gray square: periodic behavior,
black square: power law.}
\label{fig:enlargedConf}
\end{figure}

\subsection{Visualization of Computation Process}
The spectral analysis of the computation process in the URM revealed that the behavior of
sector is classified into four categories; null, white noise, sharp peaks, and power law.
Those are attributed to stationary, glider-passing, periodic, and complex behaviour of
the sector respectively.
Now let us apply these results to visualization of computation process.
We represent the category of sector's power spectrum by a mark.

The left of Figure~\ref{fig:enlargedConf} is an enlarged view of the area surrounded by solid line
in Figure~\ref{fig:InitConf} and the right of Figure~\ref{fig:enlargedConf} is made by assigning each
sector a mark according to the shape of its power spectrum.
The sector with power spectrum characterized by null, white noise, sharp peaks, and power
law is depicted by a blank, white square, gray square, and black square respectively.
At a glance we can see that almost all parts of the configuration of the URM does not work during the computation process.

The black squares are located in register 1, 7, and 10.
These registers are frequently rewritten during 65,536 time steps as shown in Table~\ref{table:EvoReg}.
This result is consistent with the observation that the sectors that has complex behavior in TM constructed on the
array of LIFE accompanies power law~\cite{Nina19}.

\begin{figure}
	\begin{center}
		\includegraphics[width=0.7\linewidth]{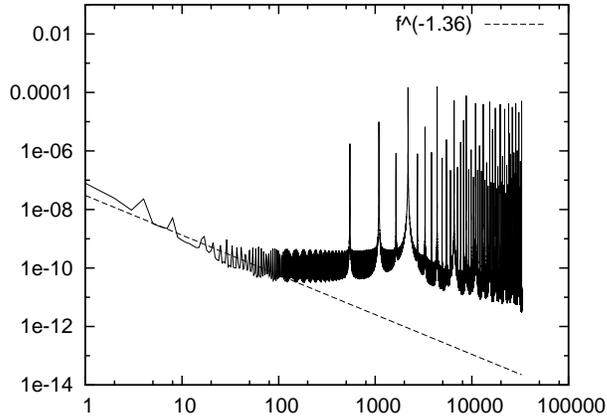}
	\end{center}
	\caption{Power spectrum averaged over the 12,160 sectors shown in the left of Figure~\ref{fig:enlargedConf}.
		The broken line represents the least square fitting of the power spectrum from $f=1$ to $f=100$ by
		by $\ln S \approx \alpha + \beta \ln f$ with $\beta = -1.36$.}
	\label{fig:wholePS}
\end{figure}

The white squares in the right of Figure~\ref{fig:enlargedConf} stand in line and
that the line represents the track of pulse moving between components.
The columns of white squares on the left end of the picture are the track of {\it latch read pulse}
and {\it latch clear pulse} moving downward from {\it latch header} to {\it latch footer}.
The track linking {\it register table} and {\it latch header} is caused by
{\it Z/NZ pulse}.
The tracks orthogonally connecting {\it register table} and {\it table area} is the trail of {\it execution pulse}
bent at {\it operation merge}.
The horizontal tracks in {\it table area} are drawn by {\it execution pulse} and {\it latch set pulse}.

\section{Discussion}
We performed regional spectral analysis of the computation process of the URM constructed on
the array of LIFE by dividing the whole array into small sectors.
The power spectrum in each sector can be classified into four categories.
The first category is trivial power spectra that have null components or have only
DC component.
This kind of power spectra means that the cell's state has never changed during the observed time steps.
The second category is white noise that is caused by rarely occurred events such as the passing of pulse.
The third category has sharp peaks caused by oscillator such as glider gun.
The fourth category is characterized by power law that is caused by complex behavior such as
frequent modification in register.
By assigning the sector a mark according to its shape of power spectrum, we can visualize
the activity of the sector during the computation process.
%This method is similar to functional neuroimaging that can visualize the local activity of brain areas.
%The neuronal activity is measured by blood flow or metabolic process in functional neuroimaging, whereas
%`computational activity' is indicated by the exponent of power spectrum in our method.

Figure~\ref{fig:wholePS} shows the power spectrum averaged over the 12,160 sectors shown in the left of Figure~\ref{fig:enlargedConf}. 
The least square fitting of the power spectrum from $f=1$ to $f=100$ by
by $\ln S \approx \alpha + \beta \ln f$ indicates $\beta = -1.36$ and  $\sigma^2 = 0.68$.
So it is considered 1/f noise in our criteria.
%It is known that Game of Life and elementary CA rule110, another computationally universal CA,
%peculiarly exhibits $1/f$ noise starting from random configuration~\cite{Nina98}, \cite{Nina08}, \cite{Nina15}.

An elementary CA rule 110 has been proved to be computationally universal through the means of emulating
cyclic tag system (CTS), another computationally universal system.
It is known that both LIFE and rule 110 have $1/f$ noise both in the transition from random configuration~\cite{Nina98}, \cite{Nina08}, \cite{Nina15}.
These results suggest that CA rules supporting computational universality bring about $1/f$ power spectrum
when it evolves from random configuration.

The evolution from random configuration exposes the genuine characteristics of the rule itself.
The behaviour during computation process is influenced not only by the rule but also by the initial configuration that is elaborately designed.
When it comes to the computation process, LIFE and rule 110 display different power spectra.
Rule 110 exhibits $1/f$ noise also in the computation process by CTS~\cite{Nina17}.
During the computation process in rule 110, the evolution goes through alternately
two phases: periodic and chaotic phase.
In the periodic phase, there never happen a collision between a stationary pattern
and a propagating pattern. 
When a propagating pattern passes through a stationary pattern, the chaotic phase
starts and lasts for a certain duration.
It seems likely that the recurrence of the periodic and chaotic phases virtually
generates intermittency.
and it is one of the mechanisms to produce $1/f$ noise~\cite{Mann80}.

On the contrary LIFE exhibits almost flat line at low frequencies in the power spectrum of the computation process
by TM~\cite{Nina19}.
%Unlike the case of CTS in one-dimensional array, Turing machine on LIFE is constructed on a two-dimensional
%array in which the signal can be detoured to avoid crossing between signals.
%So crossing signals can be avoided and that inhibit intermittency and $1/f$ noise in the computation process
%of LIFE.
The power spectrum of URM shown in Figure~\ref{fig:wholePS} is strikingly different from that of TM.
These results imply that the shape of power spectrum depends greatly on the choice of the
model of computation even though both of them are constructed on the array of LIFE.
%There are some differences in the condition between these two emulation, LIFE and URM.
%For example in the case of Turing machine emulation, the time steps observed for spectral analysis
%is $T = 16,384$, one fourth of that of URM, and the number of cells in the array is 1,430,000,
%about one thirteenth of that of URM.
%The difference in the shape of power spectrum in Turing machine and URM  

As a future plan, we are planning to coloring the sectors according the value of exponent $\beta$ and
residual sum of squares $\sigma^2$ of power spectrum. By making color image it might be able to
visualize the computational activity more vividly.

This paper is an extended version of work published in \cite{Nina19_2}.

\section{Acknowledgments}
We used the configuration of URM developed by P. Rendell downloaded from the website
\url{http://www.rendell-attic.org/gol/UCM/CMappNotes.html} and it can run on Golly, a CA
simulator.

\end{document}